\documentclass[12pt]{iopart}
\usepackage[dvips]{graphicx}

\begin{document}

\letter{Entanglement and spin
squeezing in the two-atom Dicke model}

\author{A Messikh\dag,\ Z Ficek\dag\ddag\ and M R B Wahiddin\dag}

\address{\dag\ Centre for Computational and Theoretical Sciences,
Kulliyyah of Science, International Islamic University Malaysia,
53100 Kuala Lumpur, Malaysia}

\address{\ddag\ Department of Physics, The University of Queensland,
Brisbane, QLD 4072, Australia}

\eads{\mailto{ficek@physics.uq.edu.au}, \mailto{mridza@iiu.edu.my}}

\begin{abstract}
   We analyze the relation between the entanglement and spin-squeezing
   parameter in the two-atom Dicke model and identify the source of the
   discrepancy recently reported by Banerjee and Zhou \etal that one can
   observe entanglement without spin squeezing. Our calculations
   demonstrate that
   there are two criteria for entanglement, one associated with the
   two-photon coherences that create two-photon entangled states, and
   the other associated with populations of the collective states.
   We find that the spin-squeezing parameter correctly predicts
   entanglement in the two-atom Dicke system only if it is associated
   with two-photon entangled states, but fails to predict entanglement
   when it is associated with the entangled symmetric state.
   This explicitly identifies the source of the discrepancy and explains
   why the system can be entangled without spin-squeezing. We
   illustrate these findings in three examples of the interaction of the
   system with thermal, classical squeezed vacuum and quantum squeezed
   vacuum fields.
\end{abstract}

\pacs{03.67.Mn, 42.50.Dv, 42.50.Fx}

\submitto{\JOB}

\nosections

Entanglement, the most intriguing property of multiparticle systems
(or qubits), is one of the key problems in quantum physics and has
been the subject of active research in recent years~\cite{si}.
It describes a multiparticle system which has the astonishing property
that the results of a measurement on one particle cannot be specified
independently of the results of measurements on the other particles.
Therefore, the generation of entanglement between atoms is fundamental
not only to demonstrate quantum nonlocality but also would constitute
a valuable resource in the fields of quantum information processing,
cryptography and quantum computation~\cite{ben}.
In this context, it is not surprising that a tremendous number of
theoretical proposals have been made to produce entanglement between
separate particles~\cite{ft}. Several different criteria have
been proposed to identify entanglement in two-particle systems, but no
definite measure of entanglement exists for a number of particles larger
than two. Entanglement between two particles can be identified by
calculating, for example, the Wootters entanglement measure
(concurrence)~\cite{woo}, or a measure proposed by
Peres~\cite{per} and Horodecki~\cite{horo} given in terms of the
negative eigenvalues
of the partial transposition of the density matrix of the two-particle
system. Recently, S\o rensen \etal~\cite{sdcz} have proposed a measure
of multiparticle entanglement in terms of the spin-squeezing
parameter~\cite{ku,ibbg,wbi,sor}
\begin{eqnarray}
     \xi_{\bi{n}_{i}} =\frac{N_{a}\langle \left(\Delta
     S_{\bi{n}_{i}}\right)^{2}\rangle}{ \langle S_{\bi{n}_{j}}\rangle^{2}
     +\langle S_{\bi{n}_{k}}\rangle^{2}} \ ,\label{eq1}
\end{eqnarray}
where $N_{a}$ is the number of particles, $\bi{n}_{i}, \bi{n}_{j}$
and $\bi{n}_{k}$ are
three mutually orthogonal unit vectors oriented such that the mean
value of one of the spin components, say  $\langle S_{\bi{n}_{k}}\rangle$,
is different from zero, while the other components $S_{\bi{n}_{i}}$ and
$S_{\bi{n}_{j}}$ have zero mean values. The variance
$\langle \left(\Delta S_{\bi{n}_{i}}\right)^{2}\rangle$ should be calculated
in the plane orthogonal to the mean spin direction.
A multiatom system in a coherent state has variances normal to the mean
spin direction equal to the standard quantum limit of $N_{a}/4$. In this
case, $\xi_{\bi{n}_{i}}=\xi_{\bi{n}_{j}}=1$. A system with the variance
reduced below
the standard quantum limit in one direction normal to the mean spin
direction is characterized by $\xi_{\bi{n}_{i}}<1$, that is spin squeezed
in the direction $\bi{n}_{i}$.
S\o rensen \etal~\cite{sdcz} have shown that multiparticle spin
squeezed systems also exhibit entanglement.

However, in recent studies of entanglement in the two-atom Dicke
system~\cite{ban,zsl} it has been discovered that the spin-squeezing
parameter $\xi_{\bi{n}_{i}}$ is not sufficient for predicting entanglement
in a multiparticle system. Banerjee~\cite{ban} and Zhou~\etal~\cite{zsl}
have shown that the two-atom Dicke system driven by a single mode thermal
field, can exhibit an entanglement and at the same time
$\xi_{\bi{n}_{i}}>1$. They have found that in the thermal field the time
evolution of the system is represented by a diagonal density matrix
\begin{eqnarray}
     \hat{\rho}(t) = \rho_{gg}(t)|g\rangle \langle g|
     +\rho_{ee}(t)|e\rangle \langle e|
     +\rho_{ss}(t)|s\rangle \langle s| \ ,\label{eq2}
\end{eqnarray}
where
\begin{eqnarray}
     |g\rangle &=& |g_{1}\rangle |g_{2}\rangle \ ,\nonumber \\
     |e\rangle &=& |e_{1}\rangle |e_{2}\rangle \ ,\nonumber \\
     |s\rangle &=& \frac{1}{\sqrt{2}}\left(|e_{1}\rangle
     |g_{2}\rangle +|g_{1}\rangle |e_{2}\rangle \right)  \ ,\nonumber
     \\
     |a\rangle &=& \frac{1}{\sqrt{2}}\left(|e_{1}\rangle
     |g_{2}\rangle -|g_{1}\rangle |e_{2}\rangle \right)
     \label{eq3}
\end{eqnarray}
are the collective states of the two-atom system~\cite{dic}, and
$|g_{i}\rangle, |e_{i}\rangle$ are the ground and excited states of
the $i$th atom, respectively. In the Dicke system the antisymmetric
state $|a\rangle$ is completely decoupled from the remaining states,
and then the simple three-state representation of the two-atom Dicke
system can be applied with the ground product
state $|g\rangle$, the excited product state $|e\rangle$ and the
maximally entangled symmetric state $|s\rangle$.
Since the density matrix of the system is
diagonal and the symmetric state $|s\rangle$ is a maximally entangled
state, an entanglement can be produced in the Dicke system
by a suitable population of the state $|s\rangle$. This
is exactly the situation considered by Banerjee~\cite{ban} and
Zhou~\etal~\cite{zsl}.

In this letter, we clarify the discrepancy between entanglement and
the spin-squeezing parameter. The parameter $\xi_{\bi{n}_{i}}$ has been
proposed as a simple and robust method to identify entanglement of a
large number of atoms, so we believe that a detailed analysis of the
discrepancy is of general interest.
We show that in the two-atom Dicke model,
the parameter $\xi_{\bi{n}_{i}}$ correctly predicts entanglement only
if the system is in the two-photon entangled states
which are linear superpositions of the
collective ground state $|g\rangle$ and the upper state $|e\rangle$,
but fails to predict entanglement if the system is in the entangled
symmetric state $|s\rangle$.

In order to show this more quantitatively, we start from the
definition of the parameter $\xi_{\bi{n}_{i}}$, which we can write
in terms of the density matrix elements of the system as
\begin{eqnarray}
     \xi_{\bi{n}_{i}} =2\langle \left(\Delta
     S_{\bi{n}_{i}}\right)^{2}\rangle =1 +\rho_{ss} -2|\rho_{eg}|
     \cos \theta \ ,\label{eq4}
\end{eqnarray}
where $\theta$ is the angle between $\bi{n}_{i}$ and the direction
of maximum squeezing. In the derivation of \eref{eq4},
we have used the Kitagawa and Ueda's~\cite{ku} definition
of $\xi_{\bi{n}_{1}}$ in which the variance $\langle \left(\Delta
S_{\bi{n}_{i}}\right)^{2}\rangle$, calculated in the $\bi{n}_{i}$
direction, is compared to the maximum spin
$\langle S_{\bi{n}_{k}}\rangle =N_{a}/2$ in the normal $\bi{n}_{k}$
direction.
For simplicity, we have assumed that the mean spin direction coincides
with the $z$ axis and calculated the variance in the $\bi{n}_{i}$ direction
which coincides with the $x$ axis. This is not an essential feature if
the system is driven by a thermal or squeezed vacuum field, since in
this case the mean values $\langle S_{x}\rangle$ and $\langle
S_{y}\rangle$ are zero for all values of the parameters
involved~\cite{ft}. In a more general case of a coherently driven
atoms, where $\langle S_{x}\rangle$ and $\langle
S_{y}\rangle$ are different from zero, one can adjust the angle
$\theta$ such that the maximum squeezing will coincide with the
direction of the rotated nonzero spin components.

We see from \eref{eq4} that the parameter $\xi_{\bi{n}_{i}}$
depends on the population $\rho_{ss}$ of the entangled symmetric state
and the two-photon coherence $\rho_{eg}$. Hence,
spin squeezing will be produced in the direction $\theta$ when
$|\rho_{eg}|>\rho_{ss}/2$. Note that the spin-squeezing parameter
involves the two-photon coherences with no dependence on
one-photon coherences. This indicates that the spin
squeezing can only be generated by two-photon processes.
Thus, the spin squeezing is inherent multi-atom
effect arising from the collective evolution of the Dicke system.

We now determine general conditions for entanglement in the two-atom
Dicke model using the Peres-Horodecki measure of entanglement given
by the quantity~\cite{per,horo}
\begin{eqnarray}
     E = {\rm max}\left(0, -2\sum_{i}\mu_{i-}\right) \ ,\label{eq5}
\end{eqnarray}
where the sum is taken over the negative eigenvalues $\mu_{i-}$
of the partial transposition of the density matrix $\hat{\rho}$ of the
system. The value $E=1$ corresponds to maximum entanglement between
the atoms whilst $E=0$ describes completely separated atoms.

Since the generation
of the spin squeezing is independent of the one-photon coherences, we
will look into conditions for entanglement which are determined by the
population of the collective states and the two-photon coherences.
Note, that in the Dicke model, $\rho_{aa}=0$. In
this case, the density matrix of the system in the basis $\{
|e_{1},e_{2}\rangle, |e_{1},g_{2}\rangle, |g_{1},e_{2}\rangle,
|g_{1},g_{2}\rangle \}$ can be written as
\begin{eqnarray}
\hat{\rho} &=& \left(
\begin{array}{cccc}
\rho_{ee} & 0 & 0 & \rho_{eg} \\
0 & \frac{1}{2}\rho_{ss} & \frac{1}{2}\rho_{ss} & 0 \\
0 & \frac{1}{2}\rho_{ss} & \frac{1}{2}\rho_{ss} & 0 \\
\rho_{ge} & 0 & 0 & \rho_{gg}
\end{array}
\right) \ . \label{eq6}
\end{eqnarray}
Following the Peres-Horodecki criterion for entanglement, we find that
the eigenvalues of the partial transposition of $\hat{\rho}$ are
\begin{eqnarray}
     \mu_{1\pm} &=& \frac{1}{2}\rho_{ss} \pm |\rho_{eg}| \ ,\nonumber
     \\
     \mu_{2\pm} &=& \frac{1}{2}\{\left(\rho_{ee}+\rho_{gg}\right) \pm
     \left[\left(\rho_{ee}-\rho_{gg}\right)^{2}
     +\rho_{ss}^{2}\right]^{\frac{1}{2}}\} \ .\label{eq7}
\end{eqnarray}
It is obvious that $\mu_{1+}$ and $\mu_{2+}$ are always positive. The
eigenvalues $\mu_{1-}$ and $\mu_{2-}$ become negative if and only if
\begin{eqnarray}
     |\rho_{eg}| >\frac{1}{2}\rho_{ss} \ ,\label{eq8}
\end{eqnarray}
or
\begin{eqnarray}
     \rho_{ss} > 2\sqrt{\rho_{ee}\rho_{gg}} \ .\label{eq9}
\end{eqnarray}
We are now in a position to understand quantitatively the discrepancy
between entanglement and the spin squeezing parameter.
It is seen that there are {\it two} criteria for entanglement in the
two-atom Dicke model. The first criterion, \Eref{eq8}, is
associated with the two-photon coherence and population of the
symmetric state. The second criterion, \Eref{eq9}, is associated
only with the populations of the collective states. It is evident that
the criterion \eref{eq8} overlaps with the criterion for spin squeezing,
see \Eref{eq4}. Therefore, in the absence of the two-photon
coherences, the two-atom system can still be entangled, in accordance
with the criterion \eref{eq9}, but cannot exhibit spin-squeezing,
which is associated with the criterion \eref{eq8}. This explicitly
identifies the source of the discrepancy found by Banerjee~\cite{ban}
and Zhou~\etal~\cite{zsl} and explains why the two-atom Dicke system
can be entangled without spin-squeezing.

In the situations where the criterion \eref{eq8} is satisfied, there
are entangled states generated which can be found by the
diagonalization of the density matrix \eref{eq6}. We find that the
diagonalization leads to eigenstates
\begin{eqnarray}
|\Psi_{+}\rangle &=& \left[\left(\Pi_{+}-\rho_{ee}\right)|g\rangle
+\rho_{eg}|e\rangle
\right]/\left[\left(\Pi_{+}-\rho_{ee}\right)^{2}
+\left|\rho_{eg}\right|^{2}\right]^{\frac{1}{2}} \ ,\nonumber \\
|\Psi_{-}\rangle &=& \left[\rho_{ge}|g\rangle +
\left(\Pi_{-}-\rho_{gg}\right)|e\rangle
\right]/\left[\left(\Pi_{-}-\rho_{gg}\right)^{2}
+\left|\rho_{eg}\right|^{2}\right]^{\frac{1}{2}} \ ,\nonumber \\
|\Psi_{s}\rangle &=& |s\rangle \ ,\nonumber \\
|\Psi_{a}\rangle &=& |a\rangle \ ,\label{eq10}
\end{eqnarray}
with the diagonal probabilities
\begin{eqnarray}
\Pi_{+} &=& \frac{1}{2}\left(\rho_{gg}+\rho_{ee}\right)
+\frac{1}{2}\left[\left(\rho_{gg}-\rho_{ee}\right)^{2}
+4\left|\rho_{eg}\right|^{2}\right]^{\frac{1}{2}} \ ,\nonumber \\
\Pi_{-} &=& \frac{1}{2}\left(\rho_{gg}+\rho_{ee}\right)
-\frac{1}{2}\left[\left(\rho_{gg}-\rho_{ee}\right)^{2}
+4\left|\rho_{eg}\right|^{2}\right]^{\frac{1}{2}} \ ,\nonumber \\
\Pi_{s} &=& \rho_{ss} \ ,\nonumber \\
\Pi_{a} &=& 0 \ .\label{eq11}
\end{eqnarray}
It is evident from \eref{eq10}, that in the presence of the
two-photon coherence, the system evolves
into entangled states which are linear superpositions of the
collective ground state $|g\rangle$ and the upper state $|e\rangle$.
The entangled symmetric state remains unchanged in the presence of
two-photon processes. Thus, spin squeezing and entanglement created
by the two-photon coherences are both associated with the two-photon
entangled states $|\Psi_{\pm}\rangle$.

As an example to illustrate our findings, consider the two-atom Dicke
system driven by a broadband squeezed vacuum field. In the steady-state,
nonzero matrix elements are~\cite{ft,pk90}
\begin{eqnarray}
\rho_{ee} &=& \frac{N^{2}\left(2N+1\right)
-\left(2N-1\right)|M|^{2}}{\left(2N+1\right)
\left(3N^{2} +3N +1 -3|M|^{2}\right)} \ ,\nonumber \\
\rho_{ss} &=& \frac{N\left(N+1\right)
-|M|^{2}}{3N^{2} +3N +1 -3|M|^{2}} \ ,\nonumber \\
|\rho_{eg}| &=& \frac{|M|}{\left(2N+1\right)
\left(3N^{2} +3N +1 -3|M|^{2}\right)} \ ,\label{eq12}
\end{eqnarray}
where $N$ is the intensity of the squeezed field and $M$ is the
two-photon correlation function~\cite{dfs}.

For the interaction of the system with a thermal field, $M=0$, and
then using \eref{eq12} it is straightforward to prove that
both criteria \eref{eq8} and \eref{eq9} are not satisfied for all
values of $N$. Moreover, the inequality $\xi_{\bi{n}_{x}}>1$ always
holds indicating that both entanglement and spin-squeezing are not
present in the steady-state two-atom Dicke system driven by a thermal
field.

\begin{figure}[h]
\begin{center}
\includegraphics[width=10cm]{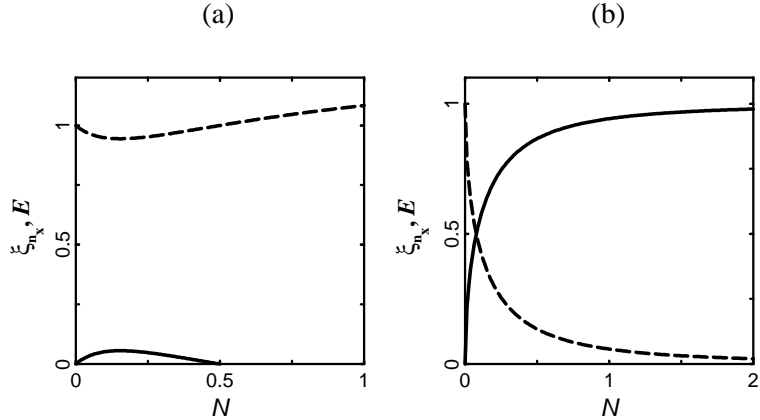}
\end{center}
\caption{\label{fig1} Entanglement measure $E$ (solid line) and the
squeezing parameter $\xi_{\bi{n}_{x}}$ (dashed line) as a function of
the intensity $N$ for (a) classical squeezed field with $|M|=N$, and
(b) quantum squeezed field with $|M|=\sqrt{N(N+1)}$. }
\end{figure}

The situation is different when the system is driven by
a classical squeezed field with the maximal two-photon
correlations $M=N$. In this case the inequality
$\rho_{ss}<2\sqrt{\rho_{ee}\rho_{gg}}$ always holds in contradiction
to \eref{eq9}. However, we find that the inequality \eref{eq8}
can be satisfied as $|\rho_{eg}|\neq 0$. According to \eref{eq4}
and \eref{eq8}, the criterion for both entanglement and spin
squeezing can be determined by positive values of a parameter
\begin{eqnarray}
     2|\rho_{eg}|-\rho_{ss} = \frac{N\left(1-2N\right)}
     {\left(2N+1\right)\left(3N +1\right)} \ .\label{eq13}
\end{eqnarray}
\Eref{eq13} shows that the system driven by the
classical squeezed field will exhibit entanglement and
spin-squeezing when $N<1/2$. We illustrate these features in
\Fref{fig1}(a), where we plot the entanglement measure $E$ and
the squeezing parameter $\xi_{\bi{n}_{x}}$
as a function of the intensity $N$. The figure clearly demonstrates
that with the condition \eref{eq8}, the squeezing parameter
correctly predicts entanglement induced by the two-photon coherences.
We should note here that the steady-state of the system driven by a
classical squeezed field is a mixed state. Thus, the squeezing
parameter correctly predicts entanglement in a mixed state if the
entanglement is generated by the two-photon coherences.

When the system is driven by a quantum squeezed field with perfect
correlations $|M|^{2}=N(N+1)$, the populations of the diagonal
states~(\ref{eq10}) are profoundly affected by the presence of the
strong two-photon correlations $M$ such that the steady-state of the
system is a pure state~\cite{ft,pk90,pa}.
Since $\rho_{ss}=0$, the criterion \eref{eq9} is not
satisfied, and therefore entanglement is determined solely by
the criterion \eref{eq8} which is always satisfied as $|\rho_{eg}|>0$.
Since the inequality $|\rho_{eg}|>0$ always holds, the system
exhibits entanglement and spin-squeezing for all $N$.
This feature is seen in \Fref{fig1}(b), where we show the
entanglement measure $E$ and squeezing parameter $\xi_{n_{x}}$ for the
quantum squeezed field. We see that the entanglement and spin
squeezing are present for all $N$. The entanglement and spin squeezing
increase with increasing $N$ and attain their maximal values, $E=1$
and $\xi_{\bi{n}_{x}}=0$, for large $N$.

As we have mentioned above, the steady-state of the system driven by
the quantum squeezed field is a pure state.
We find from \eref{eq10} that the pure state is the entangled
state given by~\cite{pk90,pa}
\begin{equation}
|\Psi_{+}\rangle = \frac{1}{\sqrt{2N+1}}
\left[\sqrt{N+1}|g\rangle
+\sqrt{N}|e\rangle\right] \ .\label{eq14}
\end{equation}
The pure state is a non-maximally entangled state, and reduces to a
maximally entangled state for $N\gg 1$.

In summary, we have clarified the discrepancy between the entanglement
and spin-squeezing parameter recently reported by Banerjee~\cite{ban}
and Zhou~\etal~\cite{zsl}. We have found that there are two criteria
for entanglement in the two-atom Dicke system, one associated with
the two-photon coherences and population of the symmetric state, and
the other associated with populations of all the collective states.
We have shown that the criterion for spin squeezing overlaps with
only one of the two criteria for entanglement, that involving the
two-photon coherences. Therefore, if entanglement is produced by the
other criterion, one obtains entanglement without spin squeezing.
Thus, our calculations demonstrate that the spin-squeezing parameter
correctly predicts entanglement in the two-atom Dicke system if the
entanglement is created by the two-photon coherences.
Moreover, the current study provides a clear physical picture of
different processes which can create entanglement in the two-atom
Dicke system.

\ack
This research has been supported in
part by Malaysia IRPA research grant 09-02-08-0203-EA002.
ZF would like to thank the International Islamic University
Malaysia for hospitality and financial support.

\end{document}